\documentclass[baaa]{baaa}

\usepackage[pdftex]{hyperref}
\usepackage{subfigure}
\usepackage{natbib}
\usepackage{helvet,soul}
\usepackage[font=small]{caption}
\pdfoutput=1

\contriblanguage{0}

\contribtype{1}

\thematicarea{9}

\title{Estudio de los efectos sistem\'aticos de SOPHIE+\\
con algoritmos de aprendizaje autom\'atico}

\titlerunning{Estudiando efectos sistem\'aticos con aprendizaje autom\'atico}

\author{
J. Serrano Bell\inst{1} \&
R.F. D\'iaz\inst{2}
}

\authorrunning{Serrano Bell \& D\'iaz}

\contact{serranojuan@fcaglp.unlp.edu.ar}

\institute{
Facultad de Ciencias Astronómicas y Geofísicas, UNLP, Argentina
\and
International Center for Advanced Studies e Instituto de Ciencias Físicas, CONICET-UNSAM, Argentina
}

\resumen{
  SOPHIE+ es un espectr\'ografo \textit{echelle} ubicado en el Observatorio de Haute-Provence, Francia. Mediante calibraci\'on simult\'anea de la longitud de onda puede alcanzar precisiones cercanas a 1~m\,s\textsuperscript{-1}. Sin embargo, el punto cero del instrumento presenta derivas a baja frecuencia de algunos ~m\,s\textsuperscript{-1} que deben ser corregidas para la alta precisi\'on que requieren los programas actuales de b\'usqueda de exoplanetas. Con este fin se monitorean regularmente cuatro estrellas de velocidad radial constante, y se usan estas mediciones para corregir las velocidades observadas. En este trabajo, proponemos una nueva forma de realizar la correcci\'on de punto cero de instrumentos como SOPHIE+. Usamos técnicas de aprendizaje automático supervisado para predecir los cambios de punto cero a partir de variables ambientales, instrumentales, y observacionales. Construimos un conjunto de datos con 645 observaciones y m\'as de 120 variables. Exploramos distintos algoritmos y logramos predecir las variaciones instrumentales de la velocidad radial con una precisi\'on de 1.47~m\,s\textsuperscript{-1}. Estas t\'ecnicas tienen el potencial de permitir realizar la correcci\'on sin necesidad de observar estrellas constantes y de obtener conocimiento sobre el instrumento que permita mejorar su estabilidad y precisi\'on.
}

\abstract{
SOPHIE+ is a echelle spectrograph located in Haute-Provence Observatory, France. It can reach a precision of near 1~m\,s\textsuperscript{-1} by simultaneus calibration. However, the zero point shows a low frequency drift of a few ~m\,s\textsuperscript{-1} that must be corrected to achieve the needed precision for the current exoplanet search programs. To this end, four radial velocity standard stars are monitored regularly to measure the instrumental drift. In this work, we propose a new way to correct the instrumental drift of instruments like SOPHIE+. We use supervised machine learning techniques to predict the zero point drift with environmental, instrumental and observational features as input. A dataset with 645 observations and more than 120 features was built. We explored various algorithms and achieved a precision of 1.47~m\,s\textsuperscript{-1} precision on the predictions of the instrumental drift. These techniques have the potential of allowing a method of correction without the need of monitoring standard stars and also can give us knowledge about the instrument that could be used to improve its stability and precision.
}

\keywords{instrumentation: spectrographs --- techniques: radial velocities --- methods: data analysis
}

\begin{document}

\maketitle

\section{Introducción}\label{S_intro}

Los espectrógrafos de alta resolución juegan hoy en día un rol crucial en la ciencia exoplanetaria, tanto para la detección de nuevos exoplanetas mediante el método de velocidades radiales (VR), como en la confirmación y caracterización de los detectados por fotometría, ya que nos permite calcular las masas. Para la detección de planetas dentro del dominio de las super-tierras se requiere una gran precisión en la determinación de las velocidades radiales. SOPHIE está
diseñado para lograr estas precisiones mediante el método de control del perfil instrumental ó \textit{IP control} \citep{2018ASSP...49..181F} y calibración simultánea de la longitud de onda. Si bien alcanza precisiones entre 1 y 2~m\,s\textsuperscript{-1}, el punto cero del instrumento muestra una deriva a largo plazo que actualmente se corrige mediante la observación
regular de estrellas estándar de velocidad radial \citep{2015A&A...581A..38C}. La idea de este trabajo es aprovechar la gran cantidad de observaciones de estrellas constantes con las que contamos en la base de datos de SOPHIE, ya que al tener una VR fija y bien conocida, la dispersión de las medidas está dominada por los errores sistemáticos del instrumento. La idea es tratar de modelar estos errores con el objetivo de proporcionar un nuevo método de corrección y a su vez intentar comprender las causas de los mismos.  
Para esto, nos proponemos entrenar un algoritmo de aprendizaje automático supervisado sobre un conjunto de observaciones de estrellas de VR constante para que aprenda a predecir las variaciones en las medidas obtenidas con SOPHIE. Usamos observaciones hechas entre 2018 y 2020, para las cuales recogimos una enorme cantidad de variables ambientales, observacionales e instrumentales que creemos que pueden estar contribuyendo a los errores sistemáticos pero que a priori no sabemos cuáles son las más importantes. El uso de algoritmos de aprendizaje automático es ideal para el tipo de problema que queremos resolver, ya que permiten manejar grandes cantidades de datos y en muchos casos nos permiten extraer información valiosa de ellos, es decir, podemos acceder a lo que ``aprende'' el algoritmo. Para la implementación de los modelos y el pre-procesado de los datos utilizamos el paquete scikit-learn\footnote{\url{{https://scikit-learn.org/}}} de Python.

\section{Metodología}
Recolectamos 645 observaciones de las estrellas HD 185144 (K0V), HD 89269 (G4V) y HD 9407 (G6.5V). Es importante incluir estrellas con distintos tipos espectrales para tener en cuenta el efecto de color \citep{2014A&A...569A..65B}. Para cada observación extrajimos 83 características (\textit{features}) de los encabezados de las imágenes y 30 más fueron colectadas de archivos externos correspondientes a sensores de temperatura y presión en distintas partes del instrumento.

\subsection{Preprocesado}
Una vez armado el conjunto de datos, lo primero que hicimos fue un análisis exploratorio para ver qué características estaban más correlacionadas con la velocidad radial y visualizar la distribución de cada característica que habíamos pre-seleccionado. Descartamos las características que tenían muchos valores malos o faltantes, las que presentaban pocos valores faltantes las completamos mediante el uso de la función \textit{SimpleImputer} reemplazando los valores indeseados por la mediana. También identificamos una pequeña cantidad de valores atípicos de las velocidades radiales y eliminamos esas filas del conjunto. Definimos nuevas características de interés a partir de combinaciones de las anteriores.

\subsection{Preparación de los datos}
\begin{figure}[!t]
\centering
\includegraphics[width=\columnwidth]{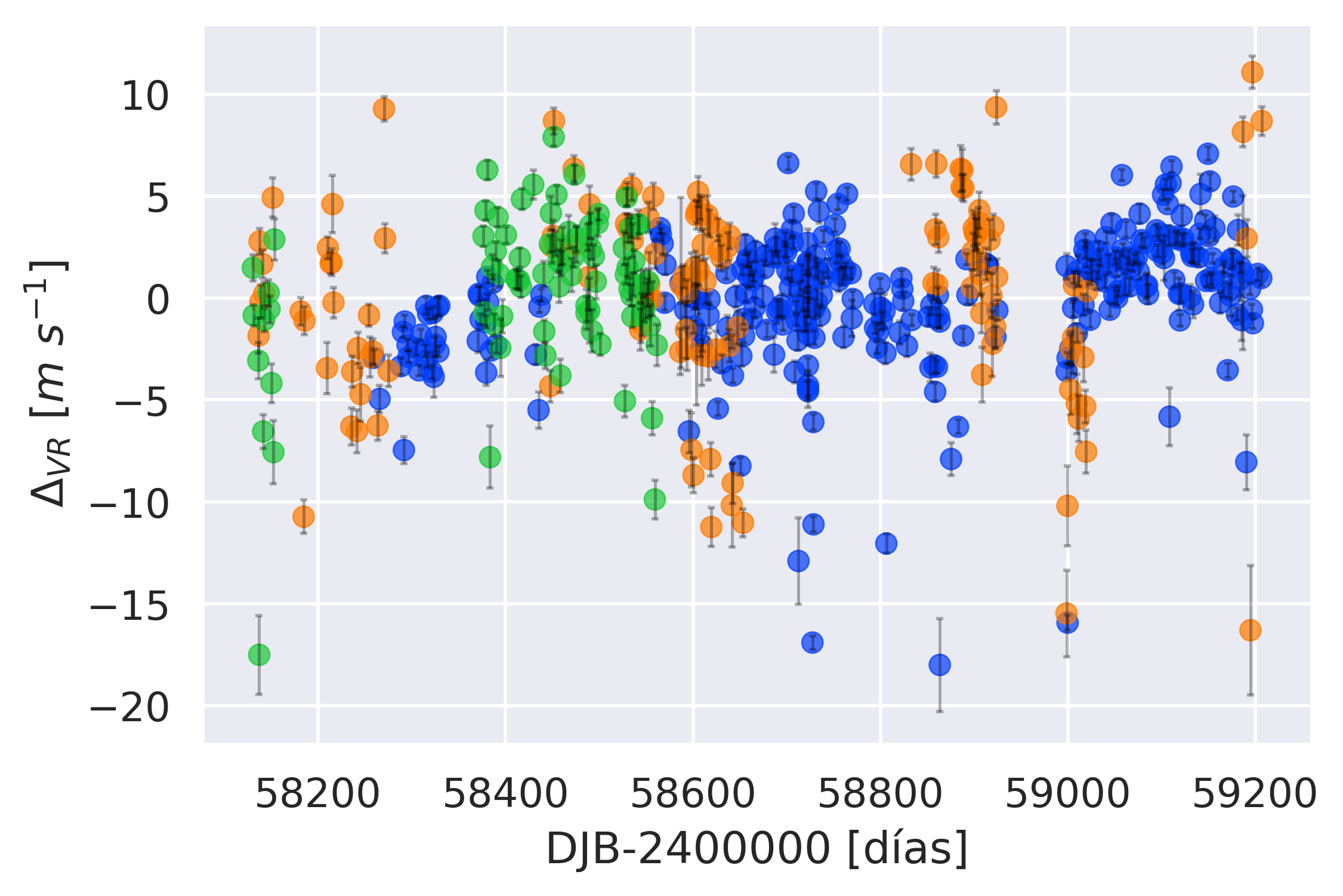}
\caption{Datos para entrenamiento por estrella. HD 185144 (azul), HD 89269 (naranja), HD 9407 (verde).}
\label{Figura 4}
\end{figure}

\begin{figure}[!t]
\centering
\includegraphics[width=\columnwidth]{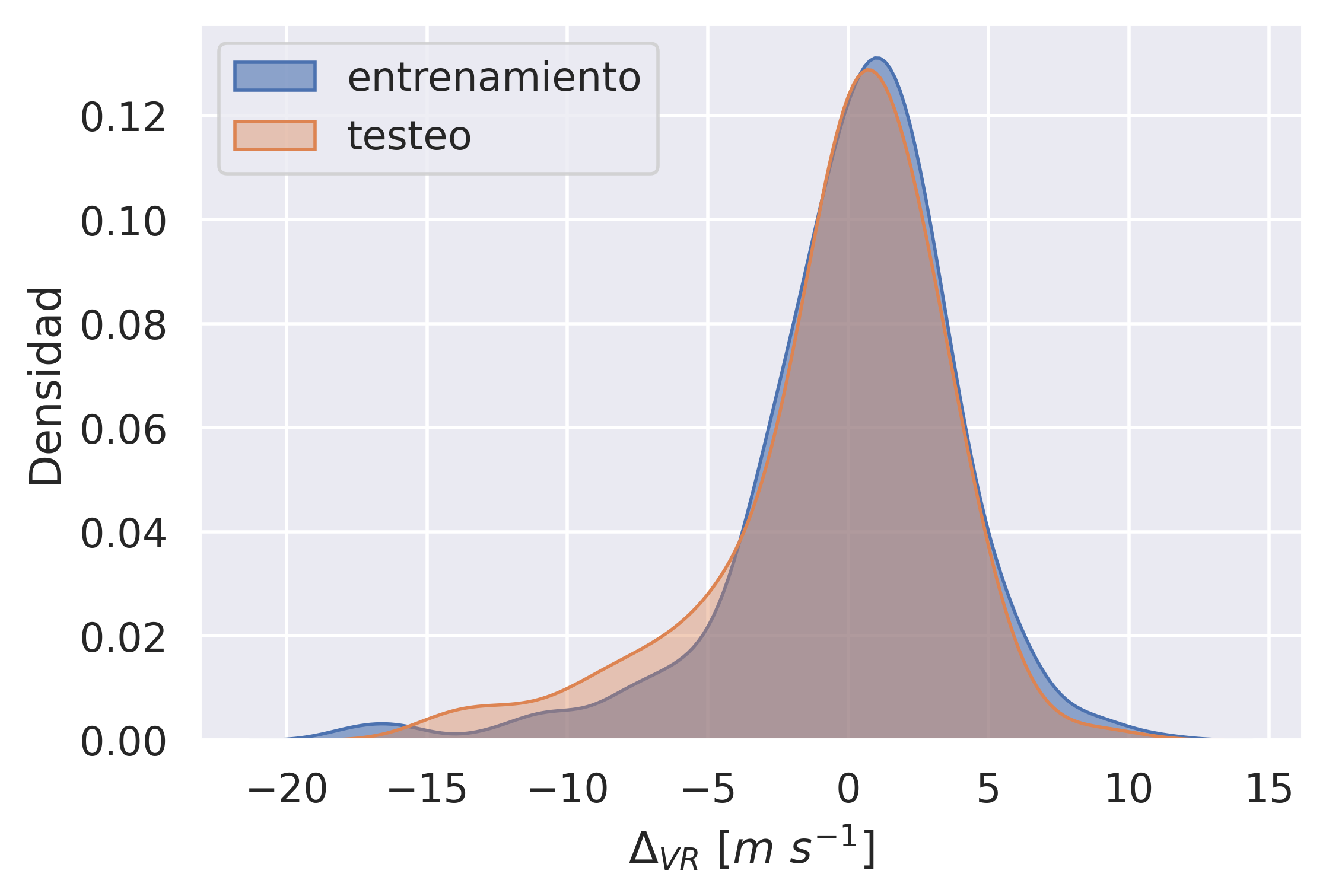}
\caption{Distribuciones de los conjuntos de entrenamiento y testeo.}
\label{Figura 5}
\end{figure}

Usando la función \textit{StandardScaler} se transformaron todas las características a media cero y varianza unitaria. Luego separamos de manera aleatoria un 20\% de los datos para testear el modelo, mientras que el otro 80\% se usó para entrenamiento de los algoritmos (figuras \ref{Figura 4} y \ref{Figura 5}).


\subsection{Entrenamiento y ajuste fino}
Probamos nueve algoritmos distintos de aprendizaje automático supervisado, entrenamos y buscamos los mejores hiperparámetros mediante validación cruzada. El algoritmo que elegimos finalmente fue un regresor lineal \textit{Lasso} ajustado con el algoritmo \textit{LARS}. Se implementó en scikit-Learn con la función \textit{LassoLars} sobre 112 características y todas las combinaciones posibles de sus productos utilizando la función \textit{PolynomialFeatures} de scikit-learn, es decir un total de 6329 características.

\subsection{Selección de características}
El algoritmo \textit{LassoLars} nos permite ver el coeficiente que tiene asociada cada característica en el modelo ya entrenado, aquí observamos que la gran mayoría eran nulos, y con esto hicimos una selección de las características más relevantes para la predicción, reduciendo la cantidad de características de 112 a sólo 33 sin afectar la precisión del modelo.

\section{Resultados}
\begin{figure}[!h]
\centering
\includegraphics[width=\columnwidth]{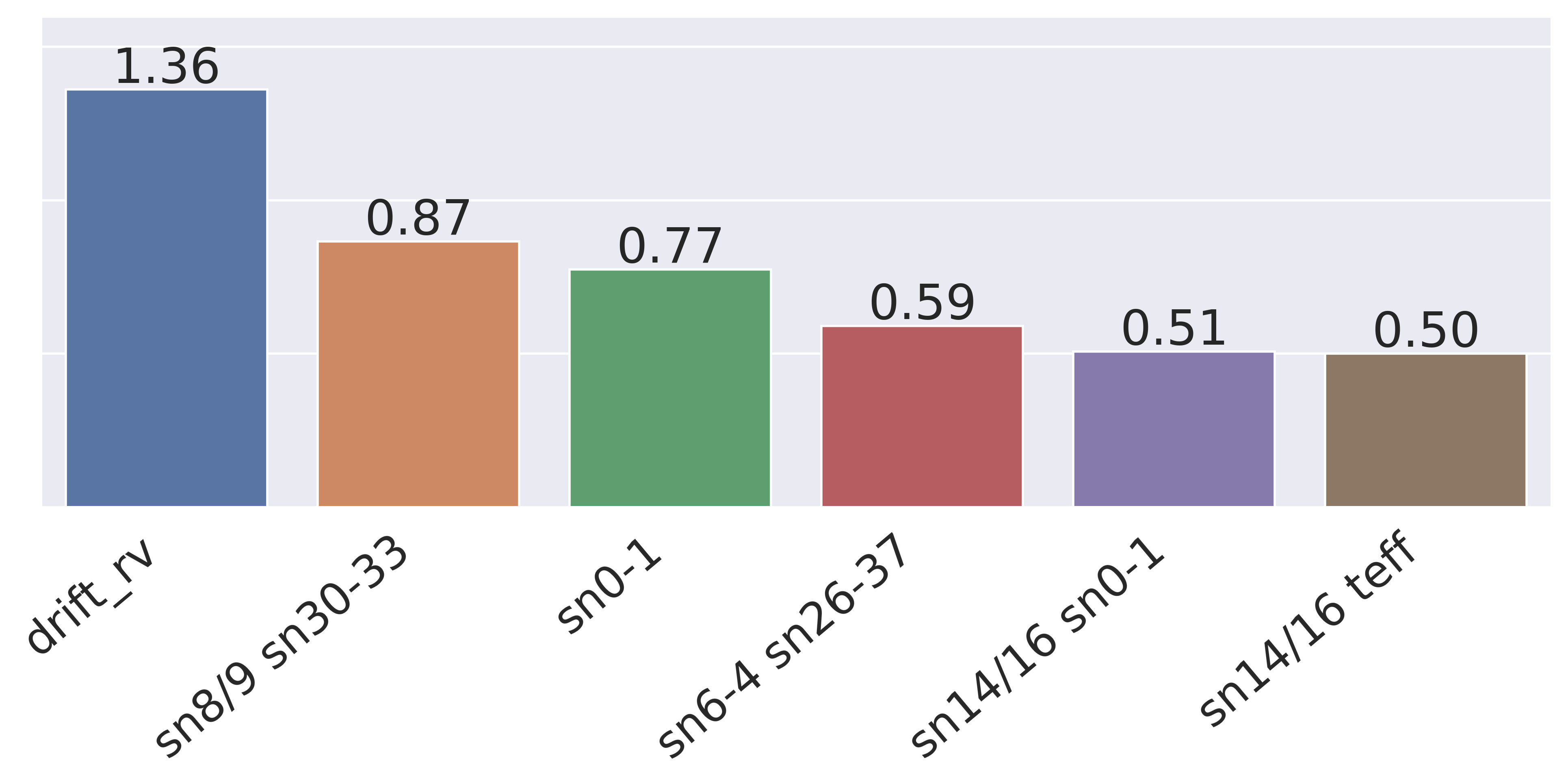}
\caption{Se muestran las seis características con coeficientes más altos en el modelo entrenado de LassoLars, los coeficientes están multiplicados por 1000. Las ``sn0'', ``sn1'', ... , ``sn\textit{i}'' son las relaciones señal a ruido del orden \textit{i} del espectro. Mientras que ``teff'' es la temperatura efectiva de la estrella. La característica más relevante es el ``drift rv'' y la segunda es el producto del cociente entre ``sn8'' y ``sn9'' con la diferencia entre ``sn30'' y ``sn33''.}
\label{Figura9}
\end{figure}

Como métrica para evaluar el modelo en el conjunto de testeo usamos el error cuadrático medio pesado (WRMSE), el cual se define como:
\begin{equation*}
    \text{WRMSE} = \sum_{i=1}^{N} \frac{w_{i}(x_{i}-\bar{x})^{2}}{wN}
\end{equation*}
donde, 
\begin{equation*}
    w_i = \frac{1}{\sigma_i^2 f_j}
\end{equation*}
es el peso del dato \textit{i}, $\sigma_i$ es la incerteza en la VR y $f_j$ es la fracción de datos del total que corresponden a la estrella $j$.
En la tabla \ref{tabla1} mostramos la desviación estándar de los datos de testeo, el WRMSE y el coeficiente R\textsuperscript{2} de las predicciones en el conjunto de testeo completo y para cada estrella por separado. Pudimos identificar a la característica llamada ``\textit{drift rv}'' como la de mayor importancia para el algoritmo (figura \ref{Figura9}), este es un parámetro que se obtiene al medir cuánto se movió el punto cero de la longitud de onda correspondiente a la lámpara de Th del instrumento desde la anterior calibración al momento de la observación.

\begin{table}[!h]
\centering
\begin{tabular}{lccc}
\hline\hline\noalign{\smallskip}
\!\!\! & \!\!\!\! Desviación & \!\!\! WRMSE & \!\!\! R\textsuperscript{2}\!\!\!\\
& \!\!\!\!Estándar [~m\,s\textsuperscript{-1}] &  [~m\,s\textsuperscript{-1}] \!\!\!\! &\\
\hline\noalign{\smallskip}
Todo  & 3.96 & 1.49 & 0.77 \\
HD 185144 & 2.63 & 1.33 & 0.67 \\
HD 89269 & 5.62 & 1.86 & 0.86 \\
HD 9407 & 4.31 & 1.67 & 0.79 \\
\hline
\end{tabular}
\caption{Evaluación del modelo en el set de testeo  completo y separado por estrella.}
\label{tabla1}
\end{table}

\begin{figure}[!h]
\centering
\includegraphics[width=\columnwidth]{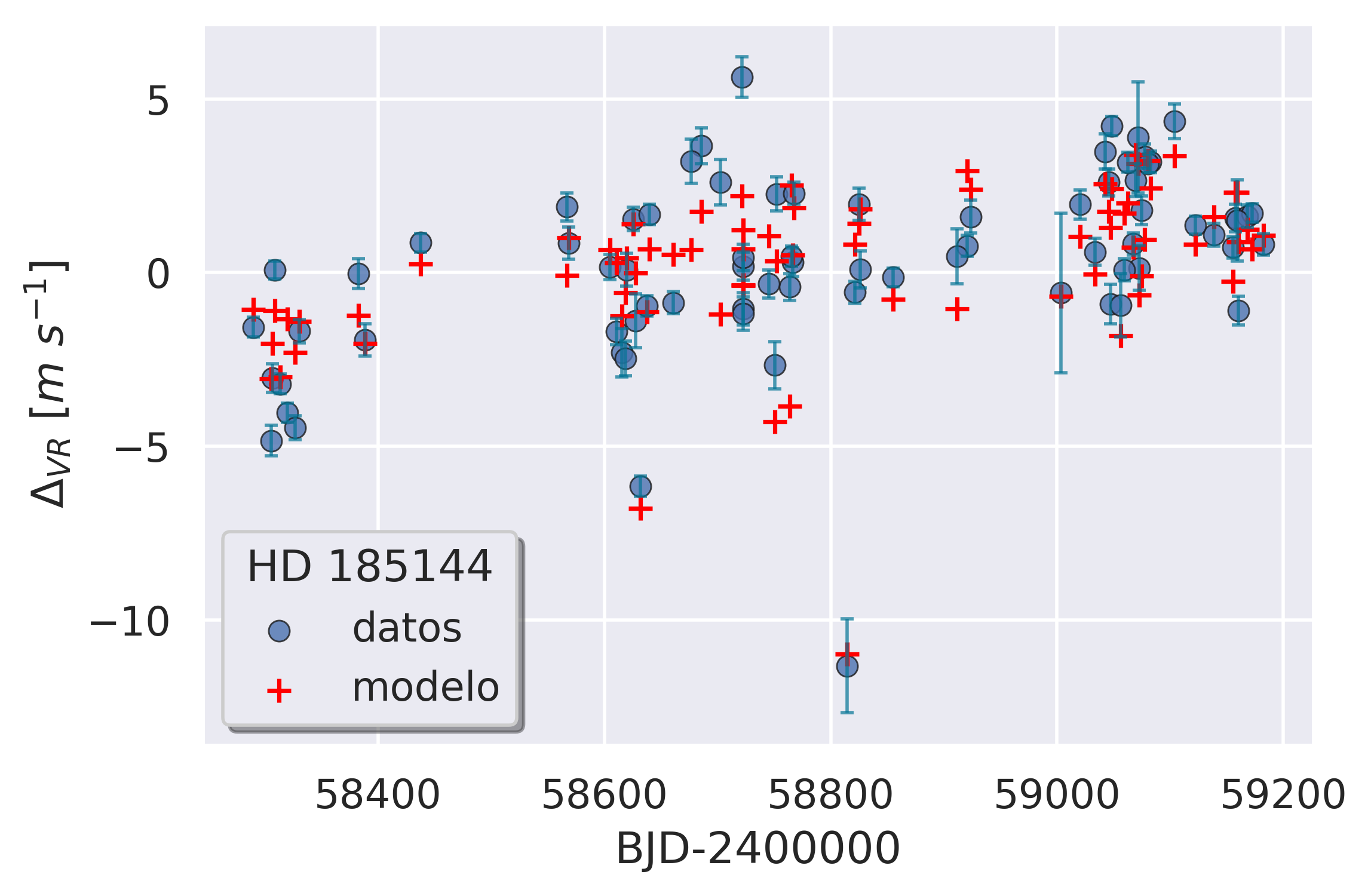}
\caption{Predicciones para HD 185144.}
\label{Figura 6}
\end{figure}

\begin{figure}[!h]
\centering
\includegraphics[width=\columnwidth]{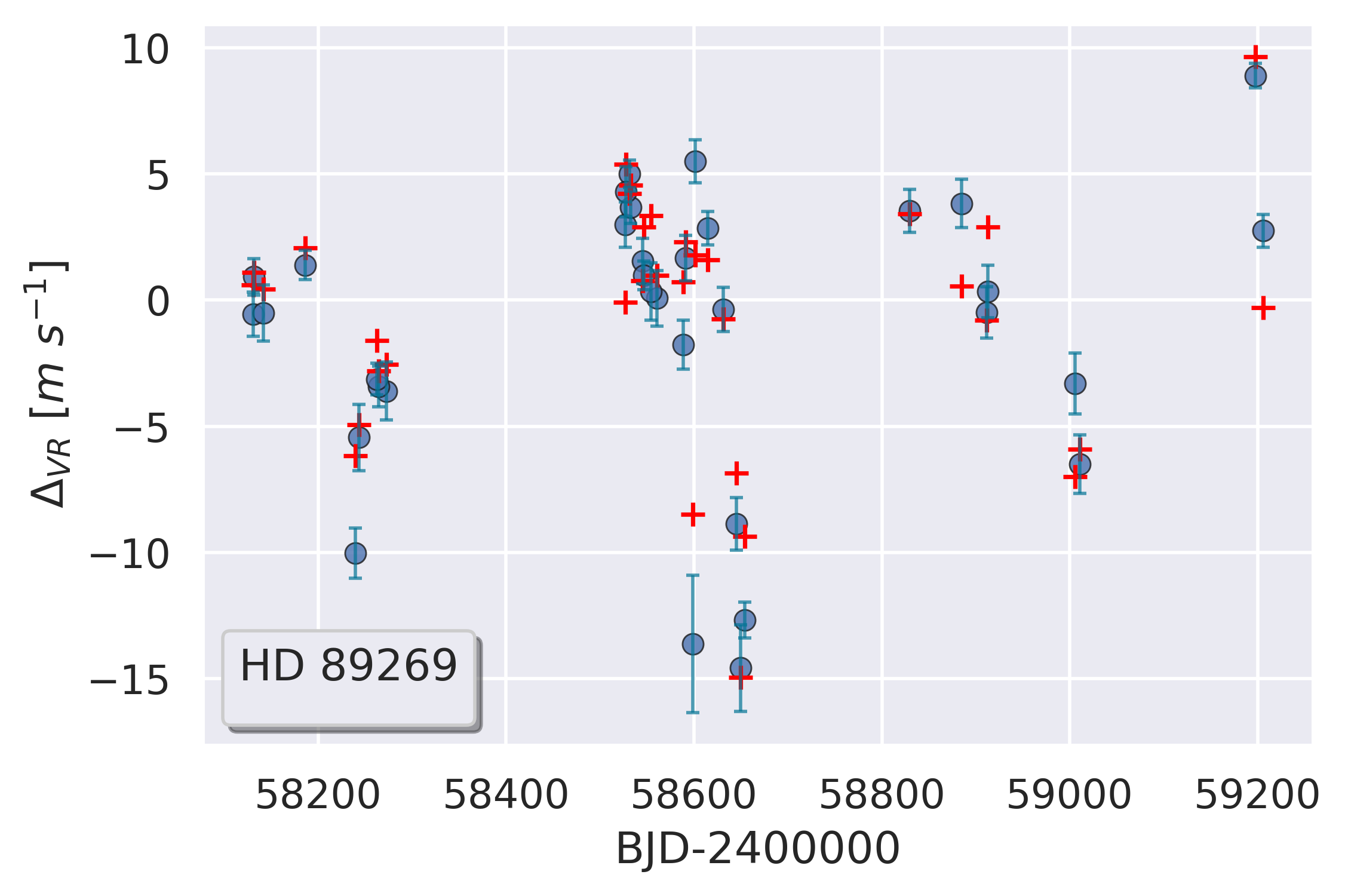}
\caption{Predicciones para HD 89269.}
\label{Figura 7}
\end{figure}

\begin{figure}[!h]
\centering
\includegraphics[width=\columnwidth]{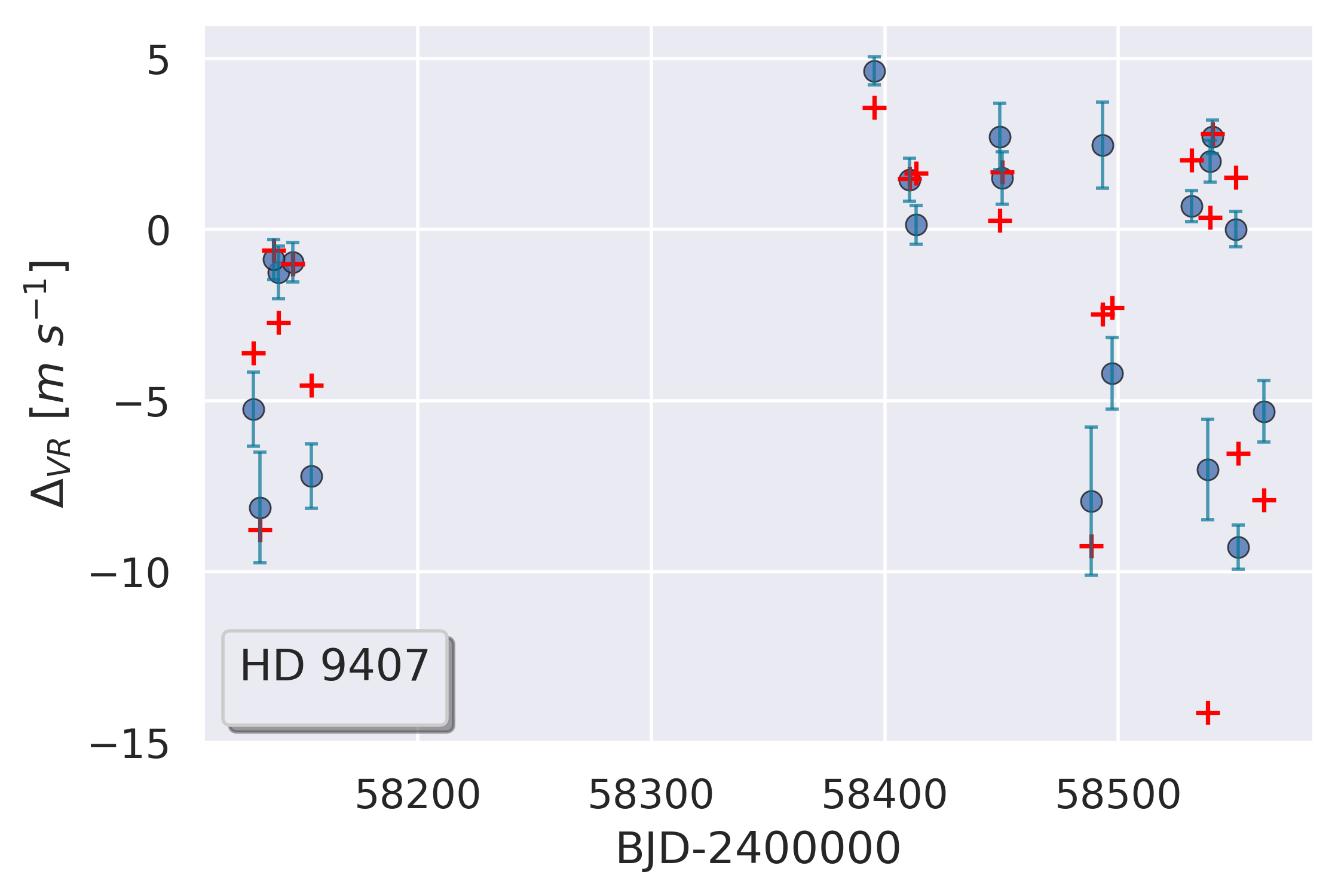}
\caption{Predicciones para HD 9407.}
\label{Figura 8}
\end{figure}

En las figuras \ref{Figura 6}, \ref{Figura 7} y \ref{Figura 8} se muestran los datos de testeo y las predicciones para cada una de las estrellas. En HD 89269 y HD 9407 vemos que el modelo explica el 86 y 79\% de la dispersión de los datos respectivamente, mientras que en HD 185144 cerca del 67\%. Para este último caso sin embargo, si aplicamos nuestro modelo como corrección a los datos de testeo nos permite reducir la dispersión de las velocidades radiales de 2.63~m\,s\textsuperscript{-1} a 1.43~m\,s\textsuperscript{-1}, lo cuál es comparable, incluso ligeramente mejor, a lo obtenido para la misma estrella por \citep{2015A&A...581A..38C} con el método tradicional de las estrellas constantes (1.51~m\,s\textsuperscript{-1}).

\section{Conclusiones y trabajo futuro}
Pudimos entrenar un modelo que predice muy bien las variaciones en las VR debida a errores sistemáticos de SOPHIE en tres estrellas de constantes y de distintos tipos espectrales. Al menos un 77\% de la dispersión de las VR puede explicarse con el modelo. El algoritmo final es un LassoLars que utiliza 33 características de las cuales identificamos a “drift rv” como la más importante, seguida de diferentes combinaciones entre las relaciones señal a ruido de los órdenes espectrales. Como trabajo futuro vamos a probar el modelo en otra estrella que no haya sido parte del conjunto de entrenamiento y en datos nuevos de las mismas estrellas.


\bibliographystyle{baaa}
\small
\bibliography{bibliografia}

\end{document}